# Quantum computers can search rapidly by using almost any transformation

Lov K. Grover, 3C-404A Bell Labs, 600 Mountain Avenue, Murray Hill NJ 07974 *(lkgrover@bell-labs.com)*

## Summary

A quantum computer has a clear advantage over a classical computer for exhaustive search. The quantum mechanical algorithm for exhaustive search was originally derived by using subtle properties of a particular quantum mechanical operation called the Walsh-Hadamard (W-H) transform. This paper shows that this algorithm can be implemented by replacing the W-H transform by almost any quantum mechanical operation. This leads to several new applications where it improves the number of steps by a square-root. It also broadens the scope for implementation since it demonstrates quantum mechanical algorithms that can readily adapt to available technology.

**0. Introduction** Quantum mechanical systems can be in a superposition of computational states and hence simultaneously carry out multiple computations in the same computer. In the last few years there has been extensive research on how to use this quantum parallelism to carry out meaningful computations. In any quantum mechanical computation the system is initialized to a state that is easy to prepare and caused to evolve unitarily, the answer to the computational problem is deduced by a final measurement that projects the system onto a unique state. The amplitude (and hence probability) of reaching a specified final state depends on the interference of all paths that take it from the initial to the final state - the system is thus very sensitive to any magnitude or phase disturbances that affect any of the paths leading to the desired final state. As a result quantum mechanical algorithms are very delicate and it is believed an actual implementation would need elaborate procedures for correcting errors [Err].

This paper shows that the quantum search algorithm is surprisingly robust to certain kinds of perturbations. It was originally derived by using the W-H transform and appeared to be a consequence of the special properties of this transform, this paper shows that similar results are obtained by substituting *any* unitary transformation in place of the W-H transform. Since all quantum mechanical operations are unitary, this means that *any* quantum mechanical system can be used - all that is needed is a valid quantum mechanical operation and a way of selectively inverting the phase of states. Meaningful computation can hence be carried out on the basis of universal properties of quantum mechanical operations, this is somewhat similar in spirit to [Neural] where circuit behavior of a certain class of neural networks was independent of the precise nature of the nonlinearity in each neuron.

**1. Quantum operations** In a quantum computer, the logic circuitry and time steps are essentially classical, only the memory *bits* that hold the variables are in quantum superpositions - these are called *qubits*. There is a set of distinguished computational states in which all the bits are definite 0s or 1s. In a quantum mechanical algorithm, the quantum computer consisting of a number of qubits, is prepared in some simple initial state, and caused to evolve unitarily for some time, and then is measured. The algorithm is the design of the unitary evolution of the system. Opera-tions that can be carried out in a controlled way are unitary operations that act on a small number of qubits in each step. Two elementary unitary operations presented in this section are the W-H transformation and the selective inversion of the amplitudes of certain states.

A basic operation in quantum computing is the operation $M$ performed on a single qubit - this is represented by the following matrix: $M \equiv \frac{1}{\sqrt{2}}\begin{bmatrix} 1 & 1 \\ 1 & -1 \end{bmatrix}$ - the state 0 is transformed into a superposition: $\left(\frac{1}{\sqrt{2}}, \frac{1}{\sqrt{2}}\right)$. Similarly state 1 is transformed into the superposition $\left(\frac{1}{\sqrt{2}}, -\frac{1}{\sqrt{2}}\right)$. In a system in which the states are described by $n$ qubits (it has $N = 2^n$ possible states) we can perform the transformation $M$ on each qubit independently in sequence thus changing the state of the system. The state transition matrix representing this operation will be of dimension $2^n \times 2^n$. Consider a case when the starting state is one of the $2^n$ basis states, i.e. a state described by a general string of $n$ binary digits composed of some 0s and some 1s. The result of performing the transformation $M$ on each qubit will be a superposition of states consisting of all possible $n$ bit binary strings with amplitude of each state being $\pm 2^{-\frac{n}{2}}$. This transformation is referred to as the W-H transformation [DJ]. This operation (or a closely related operation called the Fourier Transformation [Factor]) is one of the things that makes quantum mechanical algorithms more powerful than classical algorithms and forms the basis for most significant quantum mechanical algorithms.

The other transformation that we will need is the selective inversion of the phase of the amplitudes in certain states. Unlike the W-H transformation and other state transition matrices, the probability in each state stays the same since the square of the absolute value of the amplitude in each state stays the same. Its realization is particularly straightforward; based on [BBHT], we give such a realization.

Assume that there is a binary function $f(x)$ that is either 0 or 1. Given a superposition over states $x$, it is possible to design a quantum circuit that will selectively invert the amplitudes in all states where $f(x) = 1$. This is achieved by appending an ancilla bit, $b$ and considering the quantum circuit that transforms a state $|x, b\rangle$ into $|x, f(x) XOR\ b\rangle$ (such a circuit exists since, as proved in [Reversible], it is possible to design a quantum mechanical circuit to evaluate any function $f(x)$ that can be evaluated classically). If the bit $b$ is initially placed in a superposition $\frac{1}{\sqrt{2}}(|0\rangle - |1\rangle)$, this circuit will invert the amplitudes precisely in the states for which $f(x) = 1$, while leaving amplitudes in other states unchanged.

## 2. Amplitude amplification

A function $f(x), x = 1, 2...N$, is given which is known to be non-zero at a single value of $x$, say at $x = \tau$ - the goal is to find $\tau$. If there was no other information about $f(x)$ and one were using a classical computer, it is easy to see that on the average it would take $\frac{N}{2}$ function evaluations to solve this problem successfully. However, quantum mechanical systems can explore multiple states simultaneously and there is no clear lower bound on how fast this could be done. [BBBV] showed by using subtle arguments about unitary transforms that it could not be done in fewer than $\Omega(\sqrt{N})$ steps - subsequently [Search] found an algorithm that took exactly $O(\sqrt{N})$ steps[1].

The basic idea of [Search] was to consider an $N$ state quantum mechanical system and map each value of $x$ to a basis state of the system. The system was initialized so that there was an equal amplitude in each basis state, then by a series of unitary operations, the amplitude in the state corresponding to $x = \tau$ is increased (this state is denoted as the $|\tau\rangle$ state). A measurement is then made due to which the system collapses to a basis state, the observed basis state then gives the solution to the problem with a high probability. This algorithm was based on subtle properties of the W-H transform. The analysis in this section shows that very similar results are obtained by replacing the W-H transform by any arbitrary unitary operation. Some of the consequences of this are presented in the next section.

Assume that we have at our disposal a unitary operation $U$ and we start with the system in a basis state that is easy to prepare, say $|\gamma\rangle$. If we apply $U$ to $|\gamma\rangle$, the amplitude of reaching state $\tau$ is $U_{\tau\gamma}$, if we were to observe the system at this point, the probability of getting the right state would be $|U_{\tau\gamma}|^2$ - according to the notation, $\gamma$ denotes the initial state and $|\tau\rangle$ the target state. It would therefore take $\Omega\left(\frac{1}{|U_{\tau\gamma}|^2}\right)$ repetitions of this experiment before a single success. This section shows how it is possible to reach state $\tau$ in only $O\left(\frac{1}{|U_{\tau\gamma}|}\right)$ steps. This leads to a sizable improvement in the number of steps if $|U_{\tau\gamma}| \ll 1$.

Denote the unitary operation that inverts the amplitude in a single basis state $x$ by $I_x$. In matrix notation this is the diagonal matrix with all diagonal terms equal to $1$, except the $xx$ term which is $-1$ - a quantum mechanical implementation of this was presented at the end of section 1.

Consider the following unitary operator: $Q \equiv -I_\gamma U^{-1} I_\tau U$ - note that since $U$ is unitary, $U^{-1}$ is equal to the *adjoint* (the complex conjugate of the transpose) of $U$. We first show that $Q$ preserves the two dimensional vector space spanned by: $|\gamma\rangle$ and $U^{-1}|\tau\rangle$ (note that in the situation of interest, when $U_{\tau\gamma}$ is small, these two vectors are almost orthogonal).

---

1. $O(f(x))$ means asymptotically *less* than a constant times $f(x)$,
   $\Omega(f(x))$ means asymptotically *greater* than a constant times $f(x)$.

First consider $Q|\gamma\rangle$. By the definition of $Q$, this is: $-I_\gamma U^{-1} I_\tau U|\gamma\rangle$. Note that $|x\rangle\langle x|$, where $x$ is a basis state, is an $N \times N$ square matrix all of whose terms are zero, except the $xx$ term which is 1. Therefore $I_\tau \equiv I - 2|\tau\rangle\langle\tau|$ & $I_\gamma \equiv I - 2|\gamma\rangle\langle\gamma|$, it follows:

(1) $\quad Q|\gamma\rangle = -(I - 2|\gamma\rangle\langle\gamma|)U^{-1}(I - 2|\tau\rangle\langle\tau|)U|\gamma\rangle = -(I - 2|\gamma\rangle\langle\gamma|)U^{-1}U|\gamma\rangle + 2(I - 2|\gamma\rangle\langle\gamma|)U^{-1}|\tau\rangle\langle\tau|U|\gamma\rangle$

Using the facts: $U^{-1}U = I$ and $\langle\gamma|\gamma\rangle \equiv 1$, it follows that:

(2) $\quad Q|\gamma\rangle = |\gamma\rangle + 2(I - 2|\gamma\rangle\langle\gamma|)U^{-1}(|\tau\rangle\langle\tau|)U|\gamma\rangle$.

Simplifying the second term of (2) by the following identities: $\langle\tau|U|\gamma\rangle \equiv U_{\tau\gamma}$ & $\langle\gamma|U^{-1}|\tau\rangle \equiv U^*_{\tau\gamma}$ (as mentioned previously, $U$ is unitary and so $U^{-1}$ is equal to the complex conjugate of its transpose):

(3) $\quad Q|\gamma\rangle = |\gamma\rangle(1 - 4|U_{\tau\gamma}|^2) + 2U_{\tau\gamma}(U^{-1}|\tau\rangle)$

Next consider the action of the operator $Q$ on the vector $U^{-1}|\tau\rangle$. Using the definition of $Q$ (i.e. $Q \equiv -I_\gamma U^{-1} I_\tau U$) and carrying out the algebra as in the computation of $Q|\gamma\rangle$ above, this yields:

(4) $\quad Q(U^{-1}|\tau\rangle) \equiv -I_\gamma U^{-1} I_\tau U(U^{-1}|\tau\rangle) = -I_\gamma U^{-1} I_\tau|\tau\rangle = I_\gamma U^{-1}|\tau\rangle$.

Writing $I_\gamma$ as $I_\gamma \equiv I - 2|\gamma\rangle\langle\gamma|$ and as in (3), $\langle\gamma|U^{-1}|\tau\rangle \equiv U^*_{\tau\gamma}$:

(5) $\quad Q(U^{-1}|\tau\rangle) = U^{-1}|\tau\rangle + |\gamma\rangle\langle\gamma|(U^{-1}|\tau\rangle) = U^{-1}|\tau\rangle - 2U^*_{\tau\gamma}|\gamma\rangle$.

It follows that the operator $Q$ transforms any superposition of the vectors $|\gamma\rangle$ & $U^{-1}|\tau\rangle$ into another superposition of the same two vectors, thus preserving the two dimensional vector space spanned by $|\gamma\rangle$ & $U^{-1}|\tau\rangle$. (3) & (5) may be written as:

(6) $\quad Q\begin{bmatrix}|\gamma\rangle \\ U^{-1}|\tau\rangle\end{bmatrix} = \begin{bmatrix}(1 - 4|U_{\tau\gamma}|^2) & 2U_{\tau\gamma} \\ -2U^*_{\tau\gamma} & 1\end{bmatrix}\begin{bmatrix}|\gamma\rangle \\ U^{-1}|\tau\rangle\end{bmatrix}$

This yields a process with a periodicity of $\dfrac{\pi}{2|U_{\tau\gamma}|}$ as in [BBHT]. If we start with $|\gamma\rangle$, then after $\dfrac{\pi}{4|U_{\tau\gamma}|}$ repetitions of $Q$ we get the superposition defined by $U^{-1}|\tau\rangle$. From this, with a single application of $U$, we can get $|\tau\rangle$. Therefore in $O\left(\dfrac{1}{|U_{\tau\gamma}|}\right)$ steps, we can start with $|\gamma\rangle$ and reach the target state $|\tau\rangle$ with certainty.

The above derivation easily extends to the case when the amplitudes in states, $\gamma$ & $\tau$, instead of being inverted by $I_\gamma$ & $I_\tau$, are rotated by arbitrary phases. However, the number of operations required to reach $\tau$ will be greater. Given a choice, it would be clearly better to use the inversion rather than a different phase rotation. Also the analysis can be extended to include the case where $I_\tau$ is replaced by $V^{-1}I_\tau V$, $V$ is an arbitrary unitary matrix. The analysis is the same as before but instead of the operation $U$, we will now have the operation $VU$.

## 3. Examples of quantum mechanical algorithms

The interesting feature of this paper is that $U$ can be *any* unitary transformation. Clearly, it can be used to design algorithms where $U$ is a transformation in a quantum computer - this paper give a few such applications. The $N = 2^n$ states to be searched are represented by $n$ qubits. According to the framework of section 2, a search of the $N$ states can be carried out quantum mechanically, if we have a unitary operation $U$ which has a finite amplitude $U_{\tau\gamma}$ to go from the starting state $\gamma$ to the target state $\tau$. Such a search will take $O\left(\frac{1}{|U_{\tau\gamma}|}\right)$ steps. The following analyses calculate $U_{\tau\gamma}$ and thus the number of steps required for various algorithms - (i) & (ii) use the W-H transform as $U$; (iii) uses a different unitary operation.

**(i) Exhaustive search starting from the $\bar{0}$ state** In case the starting state $\gamma$ be the $\bar{0}$ state and the unitary transformation $U$ is chosen to be $W$ (the W-H transformation as discussed in section 1), then $U_{\tau\gamma}$ for any target state $t$ is $\frac{1}{\sqrt{N}}$. Section 2 gives an algorithm requiring $O\left(\frac{1}{|U_{\tau\gamma}|}\right)$, i.e. $O(\sqrt{N})$ steps. This algorithm would carry out repeated operations of $Q$, with $U^{-1} = U = W$, $Q$ becomes $Q \equiv -I_{\bar{0}}WI_\tau W$; the operation sequence is hence: $\ldots(-I_{\bar{0}}WI_\tau W)(-I_{\bar{0}}WI_\tau W)(-I_{\bar{0}}WI_\tau W)\ldots$ By rearranging parentheses and shifting minus signs, this may be seen to consist of alternating repetitions of $-WI_{\bar{0}}W$ & $I_\tau$.

The operation sequence $-WI_{\bar{0}}W$ is simply the *inversion about average* operation [Search]. To see this, write $I_{\bar{0}}$ as $I - 2|\bar{0}\rangle\langle\bar{0}|$. Therefore for any superposition $|x\rangle$: $-WI_{\bar{0}}W|x\rangle = -W(I - 2|\bar{0}\rangle\langle\bar{0}|)W|x\rangle = -|x\rangle + 2W|\bar{0}\rangle\langle\bar{0}|W|x\rangle$. It is easily seen that $W|\bar{0}\rangle\langle\bar{0}|W|x\rangle$ is another vector each of whose components is the same and equal to $A$ where $A \equiv \frac{1}{N}\sum_{i=1}^{N}x_i$ (the average value of all components). Therefore the $i^{th}$ component of $-WI_{\bar{0}}W|x\rangle$ is simply: $(-x_i + 2A)$. This may be written as $A + (A - x_i)$, i.e. each component is as much above (below) the average as it was initially below (above) the average, which is the inversion about average [Search].

**(ii) Exhaustive search starting from an arbitrary state** For the W-H transform, $U_{\tau\gamma}$ between *any* pair of states $\gamma$ & $\tau$ is $\pm\frac{1}{\sqrt{N}}$. We can start with any state $\gamma$ and the analysis of section 2, gives us an algorithm to reach $\tau$ in $O\left(\frac{1}{|U_{\tau\gamma}|}\right)$, i.e. $O(\sqrt{N})$, iterations. Therefore instead of starting with the $\bar{0}$ state, as in (i), we could equally well start with *any* state $\gamma$, and repeatedly apply the operation sequence $Q = -I_\gamma W I_\tau W$ to obtain an equally efficient $O(\sqrt{N})$ algorithm. The dynamics is similar to (i); however, there is no longer the convenient inversion about average interpretation.

### (iii) Search when an item *near* the desired state is known

***Problem***: Assume that an *n* bit word is specified - the desired word differs from this in exactly *k* bits.

***Solution*** The effect of this constraint is to reduce the size of the solution space. One way of making use of this constraint, would be to map this to another problem which exhaustively searched the reduced space using (i) or (ii). However, such a mapping would involve additional overhead. This section presents a different approach which also carries over to more complicated situations as discussed in [Qntappl].

Instead of choosing $U$ as the W-H transform, as in (i) & (ii), in this algorithm $U$ is tailored to the problem under consideration. The starting state $\gamma$ is chosen to be the specified word. The operation $U$ consists of the transformation $\begin{bmatrix} \sqrt{1-\frac{1}{\alpha}} & \frac{1}{\sqrt{\alpha}} \\ \frac{1}{\sqrt{\alpha}} & -\sqrt{1-\frac{1}{\alpha}} \end{bmatrix}$, applied to each of the *n* qubits ($\alpha$ is a variable parameter yet to be determined) - note that if $\alpha$ is 2, we obtain the W-H transform of section 1. Calculating $U_{\tau\gamma}$ it follows that $|U_{\tau\gamma}| = \left(1-\frac{1}{\alpha}\right)^{\frac{n-k}{2}} \left(\frac{1}{\alpha}\right)^{\frac{k}{2}}$, this is maximized when $\alpha$ is chosen to be $\frac{n}{k}$; then $\log|U_{\tau\gamma}| = \frac{n}{2}\log\frac{n-k}{n} - \frac{k}{2}\log\frac{n-k}{k}$. The analysis of section 2 can now be used - as in (i) & (ii), this consists of repeating the sequence of operations $-I_\gamma U^{-1} I_\tau U$, $O\left(\frac{1}{|U_{\tau\gamma}|}\right)$ times.

The size of the space being searched in this problem is $\binom{n}{k}$ which is equal to $\frac{n!}{n-k!k!}$. Using Stirling's approximation: $\log n! \approx n\log n - n$, from this it follows that $\log\binom{n}{k} \approx n\log\frac{n}{n-k} - k\log\frac{k}{n-k}$, comparing this to the number of steps required by the algorithm, we find that the number of steps in this algorithm, as in (i) & (ii), varies approximately as the square-root of the size of the solution space being searched.

**4. General quantum mechanical algorithms** The framework described in this paper can be used to enhance the results of *any* quantum mechanical algorithm. Assume there is a quantum mechanical algorithm $Q$ due to which there is a finite amplitude $Q_{\tau\gamma}$ for transitions from the starting state $\gamma$ to the target state $\tau$. The probability of being in the state $\tau$ is hence $|Q_{\tau\gamma}|^2$ - it will therefore take $\Omega\left(\frac{1}{|Q_{\tau\gamma}|^2}\right)$ repetitions of $Q$ to get a single observation of state $\tau$. Since the quantum mechanical algorithm $Q$ is a sequence of $\eta$ elementary unitary operations: $Q_1 Q_2 \ldots Q_\eta$, it is itself a unitary transformation. Also, $Q^{-1} = Q_\eta^{-1} Q_{\eta-1}^{-1} \ldots Q_1^{-1}$, i.e. $Q^{-1}$ is given by a sequence of the adjoints of the elementary unitary operations in the opposite order and can hence be synthesized. Applying the framework of section 2, it follows that by starting with the system in the *s* state and repeating the sequence of operations: $-I_\gamma Q^{-1} I_\tau Q$, $O\left(\frac{1}{|Q_{\tau\gamma}|}\right)$ times followed by a single application of $Q$, it is possible to obtain the state $\tau$ with certainty.

**5. Sensitivity** In order to achieve isolation, quantum mechanical computers generally have to be designed to be microscopic - however it is extremely difficult to exert precise control over microscopic individual entities. As a result, a serious problem in implementing quantum mechanical computers is their extreme sensitivity to perturbations. This paper synthesizes algorithms in terms of unitary matrices - as shown in sections 3 & 4, this framework can always be specialized to a quantum computer based on qubits; however, it can also be applied directly to more physical situations, hopefully reducing the need for error correction [Err].

For example, consider a hypothetical implementation of the quantum search algorithm where the qubits are the spin states of electrons and the W-H transform is achieved by a pulsed external magnetic field. The results of section 2 & 3 tell us that it does not significantly alter the working of the algorithm if the axes of the magnets, or the periods of the pulse are slightly perturbed from what is required for the W-H transform. Any unitary transform, $U$, *close* to the W-H transform, will work provided both $U$ & $U^{-1}$ are consistently applied as specified.

**6. Limitation** As shown in [BBHT], it is possible to express several important computer science problems in such a way so that a quantum computer could solve them efficiently by an exhaustive search. Even in physics, several important problems can be looked upon as searches of domains. Many spectroscopic analyses are essentially searches - a rather dramatic example of a recent search was that for the top quark. The framework of this paper could equally well be used here. All that is needed is a means to repeatedly apply a specified Hamiltonian that produces various phase inversions and state transitions. For example, it took about $10^{12}$ repetitions of a certain experiment, consisting of interacting a proton and antiproton at high energies, to obtain 12 observations of the top quark [Quark]. Denoting the desired state with the top quark by $|\tau\rangle$ and the initial proton-antiproton state by $|\gamma\rangle$, it implies that $|U_{\tau\gamma}|^2$ is approximately $12 \times 10^{-12}$. Therefore if it were possible to apply the operation $-I_\gamma U^{-1} I_\tau U$ repetitively $m$ times, it would boost the success probability by approximately $m^2$ (assuming $m$ to be a small number), and it would take $m^2$ fewer experiments; in case it were possible to apply the operations $-I_\gamma U^{-1} I_\tau U$, about a million times, one could achieve success in a single experiment!

In principle it is possible to synthesize $U^{-1}$ for any unitary operation, $U$, since the adjoint of a unitary operator is unitary and can hence be synthesized quantum mechanically. For controlled operations on qubits, synthesizing the adjoint is no harder than synthesizing the original operation as discussed in section 4. However, the adjoint of the time evolution operation is the reversed evolution operation - this may not be easy to synthesize when the states are non-degenerate and there is significant time evolution. This is especially true if the time-evolution is due to the internal dynamics of the system. That is the main reason this procedure, at least in its present form, could not be used to isolate the top quark!

This paper shows that the W-H transform of search based algorithms can be replaced by *any* unitary operation, provided the selective inversion operation is carried out precisely. As mentioned at the end of section 2, even the selective inversion can be replaced by a selective phase shift, provided it only affected the concerned states. Also, as mentioned in section 2, the analysis stays virtually the same if $I_\tau$ is replaced by $V^{-1} I_\tau V$ with $V$ unitary.

Another limitation is that the framework of section 2 demands that $U$ & $U^{-1}$ stay the same at all time steps. What happens if there are small perturbations in these? It seems plausible that these will not create much impact if they are small and average out to zero; however, that is something still to be proved.

**7. Conclusion**  Designing a useful quantum computer has been a daunting task for two reasons. First, because the physics to implement this is different from what most known devices use and so it is not clear what its structure should be like. The second reason is that once such a computer is built, few applications for this are known where it will have a clear advantage over existing computers. This paper has given a general framework for the synthesis of a category of algorithms where the quantum computer would have an advantage. It is expected that this formalism will also be useful in the physical design of quantum computers, since it demonstrates that quantum algorithms can be implemented through general properties of unitary transformations and can thus adapt to available technology.

The advantage of the bit representation is that the number of states that can be represented is exponential in the number of bits. After the success of classical digital computers, most quantum mechanical schemes, based them-selves on qubits. It was shown that it was possible, in principle, to synthesize gates that operate on qubits [Bit]. However independent qubits are not such a natural entity in the real quantum world as bits are in the classical world. They are very delicate entities and the slightest perturbation affects the entire computation. The distinctive feature of this paper is that it is based on general properties of unitary transforms and can hence be applied to a variety of situations including, but not limited to, qubit operations.

## 8. Acknowledgments

Thanks to Norm Margolus & Charlie Bennett for spending the time and effort to comment on several versions of this paper.